\documentclass[%
 reprint,
superscriptaddress,
 amsmath,
 amssymb,
 aps,
]{revtex4-2}

\usepackage{graphicx}
\graphicspath{{Final_plots/Fig1/}{Final_plots/Fig2/}{Final_plots/Fig3/}{Final_plots/Fig4/}{Final_plots/Appendix}}
\usepackage{comment}
\usepackage{dcolumn}
\usepackage{bm}
\usepackage{hyperref}
\newcommand{\ket}[1]{\left|{#1}\right\rangle}

\usepackage{xfrac}
\usepackage{siunitx}
\usepackage[T1]{fontenc} 
\usepackage[dvipsnames]{xcolor}
\colorlet{ForestGreen}{Black}
\begin{document}

\preprint{APS/123-QED}

\title{Efficient nuclear spin - photon entanglement with optical routing}

\author{Javid Javadzade}
 \affiliation{3rd Institute of Physics, IQST, and Research Centre SCoPE, University of Stuttgart, Stuttgart, Germany}  

\author{Majid Zahedian}
 \affiliation{3rd Institute of Physics, IQST, and Research Centre SCoPE, University of Stuttgart, Stuttgart, Germany}  
 \affiliation{Institute of Quantum Electronics, ETH Z\"urich, Switzerland}


\author{Florian Kaiser}
  \affiliation{Materials Research and Technology (MRT) Department, Luxembourg Institute of Science and Technology (LIST), 4422 Belvaux, Luxembourg}
  \affiliation{University of Luxembourg, 41 rue du Brill, L-4422 Belvaux, Luxembourg}

\author{Vadim Vorobyov}
\email[]{v.vorobyov@pi3.uni-stuttgart.de}
 \affiliation{3rd Institute of Physics, IQST, and Research Centre SCoPE, University of Stuttgart, Stuttgart, Germany}  

\author{J\"org Wrachtrup}
 \affiliation{3rd Institute of Physics, IQST, and Research Centre SCoPE, University of Stuttgart, Stuttgart, Germany}
 \affiliation{Max Planck Institute for solid state physics, Stuttgart, Germany}


\begin{abstract}
Quantum networks and distributed quantum computers rely on entanglement generation between photons and long-lived quantum memories.
For large-scale architectures, one of the most crucial parameters is the efficiency at which entanglement can be created and detected. 
Here, we maximize the efficiency for the detection of hybrid entanglement between a nuclear spin qubit in diamond with a photonic time-bin qubit.
Our approach relies on an optimal implementation of the photonic qubit analyzer, for which we use a high-speed electro-optic deflector to direct photons deterministically along the optimal interferometer paths.
This way, we completely eliminate all cases in which photons are randomly lost due to the propagation in the wrong interferometer path. 
In this first demonstration experiment, we use nitrogen-vacancy center in diamond, for which we immediately demonstrate the presence of the entanglement. 
An extension to other spin-photon entanglement systems is straightforward.
Further, our scheme can be used in the framework of quantum repeater networks, including spectral and temporal multiplexing strategies. 
Our results thus pave the way for the future high-performance quantum networks.
\end{abstract}

\maketitle

\section{Introduction}
Quantum networks \cite{Kimble_2008, jiang2007distributed,childress2005fault} are emerging as a novel paradigm for the distribution of fragile entangled quantum states among individual parties, challenged by the loss in the communication channel. 
Thus, efficient distribution and detection of the quantum entanglement in such a network is a key challenge.
Among various proposals, such as a twin-field quantum key distribution \cite{lucamarini2018overcoming} or cluster states \cite{borregaard2020one}, a quantum repeater architecture consisting of a stationary and a flying qubits \cite{van_Loock_2020}  is currently the main avenue for experimental demonstrations \cite{stas2022robust, hermans2022qubit, langenfeld2021quantum, krutyanskiy2023entanglement}. 
The development of scalable and efficient quantum communication link will open new possible applications, such as secure communications \cite{ekert1991quantum}, distributed quantum computing \cite{de2024thresholds}, clock synchronisation \cite{shi2022clock}, long range terrestrial based optical telescopes \cite{khabiboulline2019optical} or distributed quantum sensing and metrology  \cite{zhao2021field}.
When comparing protocols, one finds schemes based on photon emission and absorption \cite{van_Loock_2020}. 
Absorption based schemes do not have a time overhead related to the time of flight of the photons but are limited in efficiencies in the absence of a good spin-photon interface \cite{van_Loock_2020}. 
In this work, we consider an emission based scheme, which is favourable for systems like neutral atoms and colour centers at the current level of technology \cite{rozpkedek2019near}. 
An important ingredient of the scheme is a quantum memory, which stores the entangled qubit state during the time required for the photon to propagate to the measurement station. 
Typically for a \SI{100}{km} inter-repeater distance, it takes light  $\tau \sim \,\SI{0.5}{ms}$ to travel between the stations. 
For a robust operation, the coherence time should be significantly higher to account for probabilistic heralding efficiencies \cite{van_Loock_2020}. 
To this end, a robust nuclear spin memory is required, to store the entanglement, during the photon time of flight, and potentially for establishing an entanglement with the second link in case of a quantum repeater \textit{segment} operation mode \cite{rozpkedek2019near}.
\begin{figure*}
	\centering
	\includegraphics{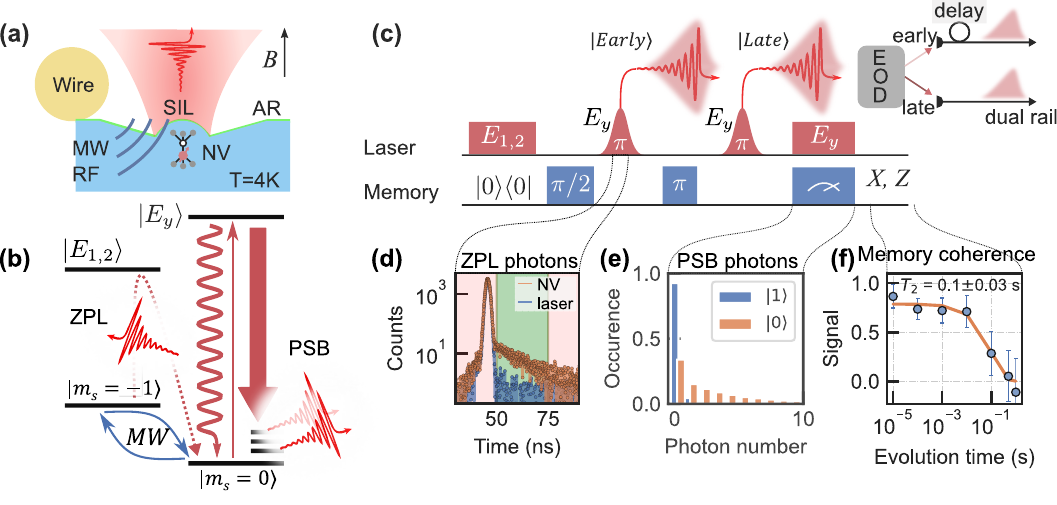}
	\caption{ Experimental setup. \textbf{(a)} Sample layout with diamond containing an NV-center in a solid immerion lens in a cryostat at T=4K. MW and RF are supplied via a \SI{20}{\mu m} wire. The light from the NV center is collected with high NA objective. \textbf{(b)} Energy level diagram. The $E_y$ transition was used for spin-photon entanglement and readout, the $E_{1,2}$ transition was used for spin initialization, MW and RF fields were used to manipulate electron and nuclear subleves for coherent control of the memory. 
	\textbf{(c)}. General protocol scheme. The laser fields initialise spin, following by preparation of a coherent superposition. Optical $\pi$ pulse applied to $m_s = 0$ sublevels is used for the creation of the spin-photon entanglement. Between two optical $\pi$-pulses MW $\pi$ pulse was used to exchange populations in the ground state qubit state. 
\textbf{(d)} The zero photon line photons temporal shape and comparison to the $E_y$ laser induced background \textbf{(e)} The single shot readout phonon sideband photon counting statistics for readout of the memory state \textbf{(f)} Nuclear spin memory coherence time measurement with a Hahn echo type protocol.}
	\label{fig1}
\end{figure*}
Various platforms, e.g. trapped atoms, rare earth ions and color centers, are suitable for the role of optically accessible quantum memories capable of generating entangled memory-photon systems \cite{azuma2023quantum}.
In this work, we employ nitrogen vacancy (NV) center in diamond, a well-established and understood model system, as a working platform \cite{doherty2013nitrogen}. 
A robust to optical path fluctuation and compatible with optical memory encoding of the photonic mode is essential for operation of the repeater link.
In the past it was shown that time-bin schemes are particularly suitable for longer optical distances, as they are tolerant against optical path length fluctuations \cite{knaut2023entanglement, beukers2024remote}, polarization drifts and allows for a straghtforward implementation of laser suppression schemes \cite{kuhlmann2013dark,benelajla2021physical}. 
However, conversion of this encoding into interferometer relevant dual rail encoding is done via unbalanced interferometers pathways. 
This leads to a typical loss of 50\% per node of the usable photons and reduces the rate of entanglement distribution significantly.
For a few nodes  this is still afordable, however it scales unfavourable when it comes to linking multiplexed repeater nodes.
We address this issue by using an optical switch for routing single photons to the relevant branches of the interferometer, which yields nearly 100\% efficiency of conversion.
Besides this we combine all elements of an efficient repeater segment, namely a solid state quantum memory exceeding hach echo coherence time $T_{2}^{HE} \sim \SI{100} {ms}$, an optically coherent spin dependent emission, and an on demand fast optical switching among interferometer arms. 
With that we show an efficient entanglement generation, storage and delivery via a NV center-based spin photon interface.  
Specifically, we store entanglement in a long lived nuclear spin memory and deliver the entanglement to dual rail mode via an optical deflector integrated into the unbalanced interferometer. 
Our results bring the highly efficient quantum repeaters step closer to its optimal performance. 

\begin{figure}
	\centering
	\includegraphics{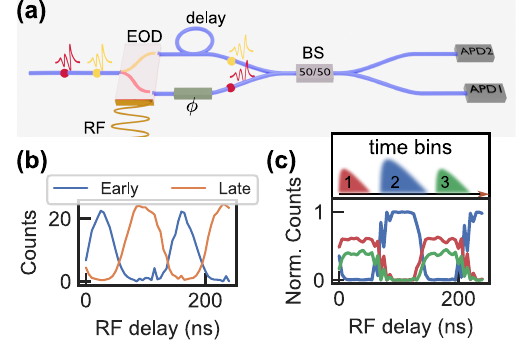}
		\caption{ \textbf{(a)} Unbalanced Optical interferrometer with electro-optical Deflector \textbf{(b)} Case of sending single photon. Depending on RF fields delay the photon is appeared in early or late mode, due to interferrometer imbalance \textbf{(c)} Case of sending a pair of early and a late photons. Depending on the RF delay, both photons either appear in middle time bin 2, erasing the early-late information, or appear in bins 1 and 3 leading to early-short and late-long correlations}
		\label{fig2}
\end{figure}

\section{Results}
The central part of our interface is an NV colour center in diamond (Fig. \ref{fig1}a) in a $T=\SI{4}{K}$ optical cryostat. 
We exploit a solid immersion lens (SIL) in $\langle 111 \rangle$ diamond, hosting a single colour center in its center. 
The lens is coated with an antireflection coating (AR) for suppression of the laser reflection from the surface and optimising light collection from the emitter.
Throughout this work we use the $E_y$ optical transition for spin readout and pulsed coherent optical excitation. The $E_{12}$ transition is used for spin initialisation into $m_s = 0$ state (Fig. \ref{fig1}b). 
We coherently drive the electron spin with microwave fields supplied via a copper wire and nuclear spin with radio frequencies supplied via the same wire. 
The magnetic field of around $B \sim \SI{80}{G}$ is aligned with the $\langle 111 \rangle$ crystal axis of the diamond and NV center symmetry axis. 
In the ground state, the NV center has an electronic spin of $S=1$.
We use the two ground state sublevels $m_s = 0  \equiv |0\rangle$ and $m_s = -1  \equiv |1\rangle$ in this work shown in Fig. \ref{fig1}b. 
Our spin-photon entanglement generation protocol is schematically shown in Fig. \ref{fig1}c, and further detailed in the following.
We start by initializing the NV center’s spin state into $|0 \rangle$ using $E_{1,2}$ laser excitation. 
Next, we create a coherent superposition state between the two relevant ground states using a MW pulse, i.e., $\psi_0 = (|0 \rangle  + |1 \rangle) / \sqrt{2}$. 
We then follow-up with an optical laser excitation pulse along the transition $E_y$. 
This can lead to a subsequent \textit{early} photon emission into the zero-phonon line, conditioned to NV center spin state $m_s = 0$. 
The intermediate state after this first optical excitation is then $\psi_1 = \left( |0 \rangle \hat{a}^{\dagger}_{\rm early}  + |1 \rangle / \sqrt{2} \right) |{\rm vac} \rangle$, where $\hat{a}^{\dagger}_{\rm early}$ is the photon creation operator that acts on the vacuum state $|{\rm vac} \rangle$.
In the next step, we use a MW $\pi$-pulse to exchange the NV center’s ground state spin populations, $|0\rangle \leftrightarrow |1 \rangle$. 
Finally, we apply a second optical laser excitation pulse along the transition $E_y$. This creates the final spin-photon correlated target state:
\begin{equation} \label{eq:psi2}
\psi_2 = \left( |1 \rangle \hat{a}^{\dagger}_{\rm early}  + |0 \rangle \hat{a}^{\dagger}_{\rm late} / \sqrt{2} \right) |{\rm vac} \rangle.
\end{equation}
 A typical single-photon decay response signal into the zero-phonon line after laser excitation on the $E_y$ transition is shown in Fig. \ref{fig1}d. 
At long time delays, it shows the typical mono-exponential lifetime response of the NV center. 
The significant peak at short delays is explained by a residual amount of excitation laser breakthrough, which we could not fully compensate in our cross-polarization setup \cite{kuhlmann2013dark, benelajla2021physical} (for more details, see SM section I).
To infer the nature of the created state, we probe the $X$ and $Z$ component of the electron spin via the microwave control pulse and single shot readout . 
A characteristic photon counting histogram is presented in Fig. \ref{fig1}e  (see \cite{zahedian2023readout} for details). 
The associated with single NV center $^{14}N$ nuclear spin presents a resource for quantum memory, exceeding storage of 100 ms (Fig. \ref{fig1}f), which will be used to store the created electron spin - photon entanglement. 
To probe the created spin-photon entangled state in the $XX$ basis, one requires a photonic qubit analyzer in a form of unbalanced Mach-Zehnder interferometer. 
It serves to map temporal modes to spatial ones, facilitating qubit operations. 
Typically, the interferometer has long and short arms with 50/50 beam-splitter at the input, reducing the efficiency of the mapping by 50 \% when an unbiased input is used.
To overcome this deficiency, we optimize temporal to spatial conversion to become essentially deterministic by using an active photonic routing device at the input of the interferometer depicted in Fig. \ref{fig2}a.
It deliberately routes the $\ket{Early}$ time bin mode to the $\ket{Long}$ spatial mode of the interferometer and $\ket{Late}$ time bin mode to the $\ket{Short}$ spatial mode, thus increasing the efficiency of the link by a factor of two and reducing the losses of the events. 
For this we use an Electro Optical Deflector (EOD) provided by QUBIG GmbH, working at a switching rate of \SI{7.14}{MHz} (\SI{70}{ns}) Fig. \ref{fig2}a. 
Photons from a single temporal mode at the input can be guided to the $\ket{Short}$ or $\ket{Long}$ branch of the interferometer, resulting in mapping into early or late temporal mode Fig. \ref{fig2}b with fidelity $\mathcal{F}_{eod}\sim97\%$. 
If the early and late photon modes are entering the interferometer,  by adjusting the deflector RF drive delay, one can achieve desired erasure of temporal information, thus mapping both temporal photon modes into spatial modes (Fig. \ref{fig2}c). 
The EOD suppresses two side peaks (1 and 3) and increases the desired efficiency of conversion from $50\%$ to $\sim 97 \%$ in this way ensuring efficient conversion from time-bin to spatial modes. (see SM section III for details)

\begin{figure*}[t]
	\centering
	\includegraphics{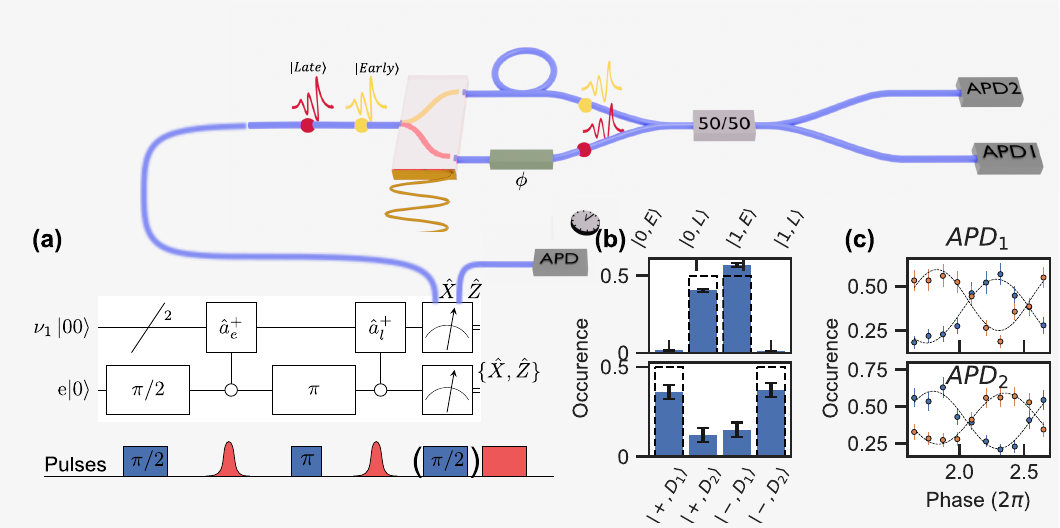}
		\caption{Electron Spin - Photon - Entanglement (SPE) \textbf{(a)} SPE sequence with quantum diagram and a photonic qubit analyzer.
		\textbf{(b)} Measured Electron spin - photon correlations in $XX$ $(0.46 \pm 0.08)$ and $ZZ$ $(0.95 \pm 0.02)$ bases, leading to a total Fidelity of $\mathcal{F}\ge 0.70 \pm 0.08 \%$ \textbf{(c)}  Probability of measuring the NV center spin in state $|0\rangle$ (blue) and $|1\rangle$ (orange) and detector APD$_1$ or APD$_2$ click depending on the phase $\phi$ of the optical interferometer.}
	\label{fig3}
\end{figure*}

We first test the full protocol on the electron spin as a memory qubit, presented in Fig. \ref{fig3}a. 
Following the protocol for creating the correlated $\psi_2$ (eq. \ref{eq:psi2}) entangled state, we further use the actively routed interferometer as a photonic qubit analyzer 
and estimate the fidelity of the generated state in XX basis. 
For a lower bound on the fidelity we measure correlations of both qubits in $X$ and $Z$ basis \cite{vasconcelos2020scalable} and use a lower bound estimation:
\begin{equation} \label{eq:2}
\ F \geq \frac{1}{2} (\rho_{22}+\rho_{33} -2 \sqrt{\rho_{11} \rho_{44}} +\tilde{\rho_{11}}+\tilde{\rho_{44}} -\tilde{\rho_{22}}-\tilde{\rho_{33}}),
\end{equation}
where $\rho_{ii}$ are the diagonal elements of the density matrix in $ZZ$ basis, and $\tilde{\rho_{ii}}$ - in $XX$ basis.
To experimentally estimate the transverse matrix elements we utilize interferrometer with controllable phase acquired by a photonic qubit converted to dual rail basis:
\begin{equation} \label{eq:before2BSstate}
\psi_3 = \frac{\ket{0,Short}+e^{i\phi} \ket{1,Long}}{\sqrt {2} }.
\end{equation}
We actively control the phase $\phi$ via a voltage controlled fiber-stretcher in one of the arms Fig. \ref{fig3}a. 
After a final beamsplitter, the probability of detecting a photon and NV center spin in particular state at the corresponding detector follows (see SM section IV):
\begin{equation}
  \begin{split}
p ( \ket{0}, APD_1) &=  \frac{1-\cos \phi}{4}, \\ 
p ( \ket{1}, APD_1) &=  \frac{1+\cos \phi}{4}, \\
p ( \ket{0}, APD_2) &=  \frac{1+\cos \phi}{4}, \\ 
p ( \ket{1}, APD_2) &=  \frac{1-\cos \phi}{4}.
  \end{split}
\label{eq:ZPLvsMsProbs}
\end{equation} 
We experimentally measure the probabilities in eq. \ref{eq:ZPLvsMsProbs}, by running the aforementioned protocol using the interferrometer with EOD. 
By using the FPGA logic, we apply a final readout of the state conditional on detecting the photons either in $APD_1$ or $APD_2$ modes, thus saving measurement time (see SM section II). 
In Fig. \ref{fig3}c one can see the probability of finding the NV center in states $\ket{0}$ and $\ket{1}$ depending on the phase of the interferometer, after heralding it with click of the photodetector $APD_1$ or $APD_2$ respectively.
Given the low probability of detecting an emitted photon, the experiment requires multiple repetitions.
This emphasises the need for stability of both the interferometer and the spectral characteristics of the emitted photons. 
To overcome this challenges we have made active optical transition stabilisation via the charge resonance check (CRC) performed with a standalone FPGA, working as a master device in the experiment  (see SM section II for technical details).
To estimate the correlations in the $ZZ$ basis, we measured probabilities to find the spin and photonic qubits simultaneously in different basis states as depicted in Fig. \ref{fig3}b (top).
Fig. \ref{fig3}b illustrates the experimental outcomes of measuring two qubits in either the $X$ or $Z$ basis. 
Specifically, it depicts the probabilities associated with measuring a two-qubit system in four distinct basis states. 
Thus, experimentally obtained values are $0.46 \pm 0.08$ for $XX$ and $0.95 \pm 0.02$ for $ZZ$ correlations measurements.
Using eq. \ref{eq:2} one can estimate the fidelity of entangled state of $\mathcal{F}\ge 0.70 \pm 0.08 \%$.
To find the $XX$ basis we perform a phase adjustment over the whole $2\pi$ range and detect characteristic sinusoidal dependence described by the eq. \ref{eq:ZPLvsMsProbs}.

After accounting for detector dark count events, we achieve an overall contrast of $\approx 62 \%$, which could be explained by imperfect initialisation and single shot readout $\sim 80  \%$, instability of the interferometer $97 \%$, misalignment of the interferometer $95 \%$ and residual infidelity of $84 \%$ we associate with defect spectral diffusion of $\sim 3$ MHz. 
We note that these numbers are not a fundamental limitation of the scheme and could be further improved on the expense of entanglement generation rate in our setup. 
The achieved result demonstrates the ability to measure the phase collected by a photon on the NV electron spin, as a nonlocal quantum sensor. 
We note that without dark count correction our fidelity was not significantly reduced due to the initially low dark counts, low strained diamond sample and excellent optical properties of the defect.

\begin{figure}[ht]
	\centering
	\includegraphics[width=\columnwidth]{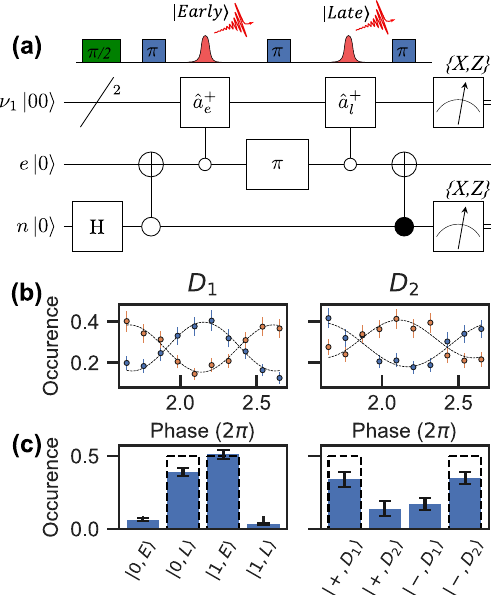}
	\caption{Nuclear Spin - Photon - Entanglement \textbf{(a)} Nuclear SPE sequence with quantum diagram. The electron spin was first entangled with the nuclear spin and then used to create entanglement with the photon. After measuring a photon, the joint (entangled) state of electron and nuclear spin was projected \textbf{(b)} Probability of measuring the nuclear spin in state $|0\rangle$ (blue) and $|1\rangle$ (orange) and detector APD$_1$ or APD$_2$ click depending on the phase $\phi$ of the optical interferometer \textbf{(c)} Dashed bars - values, corresponding to ideal case, blue bars - measured Nuclear spin-photon correlations in $XX$ ($0.38 \pm 0.09$) and $ZZ$ ($0.81 \pm 0.04$) basis, leading to a total Fidelity of $\mathcal{F}\ge 0.6 \pm 0.1$  }	
	\label{fig4}
\end{figure}

Further, we store the obtained entangled spin-photon state in a long lived quantum memory of $^{14}N$ nuclear spin.
Before running the protocol, we first initialize the nuclear and electron spins to the initial state $\ket{00}$ (see SM section V). 
As a first step we apply an RF $\pi / 2$ pulse between nuclear sub-levels to create their equal superposition $\ket{0}(\ket{0}+\ket{1})/\sqrt {2} $, while keeping the electron spin in state $\ket{m_s = 0}$. 
Then, by applying a $\pi$ rotation of the electron spin, conditioned on the nuclear spin (CNOT gate), we create an entangled state of the 2-spin system  $(\ket{10}+\ket{01})/\sqrt {2}$. 
The next step is to create an entanglement between the electron spin and the photon using the above mentioned method (Fig. \ref{fig3}a). Thus we add a third qubit (photon) to the entangled system, resulting in the nuclear spin - electron spin - photon entanglement eq.\ref{eq:3}. 
\begin{equation} \label{eq:3}
\frac{\ket{11Early}+\ket{00Long}}{\sqrt {2}}
\end{equation}
At last, the repeated application of the CNOT operator to the electron spin serves to disentangle it.
This not only holds potential for extending the spin coherence time of the nuclear spin but is also a crucial step in confirming that our measurements indeed reflect the correlation between the photon and the nuclear spin. 
Due to the fact that there is no direct access to nuclear spin state measurement, we read its state using the electron spin. 
Similarly to electron spin $X$ readout, we apply an RF $\pi/2$ rotation with following $Z$ projection readout (Fig. \ref{fig4} b).
The achieved fidelity with nuclear spin is $0.38 \pm 0.09$ $(XX)$ and $0.81 \pm 0.04$ $(ZZ)$, yielding overall fidelity to be $\mathcal{F}\ge 0.6 \pm 0.1$ and showing the potential of the system. 
As in the case of the electron-photon entanglement, the fidelity is limited by reduced readout and initialisation fidelity of the electron spin in the presence 
of cross-polarisation scheme, which can be overcome by increased thresholds in readout estimation method and subsequent reduction of measurement success rates. 
Additionally, the nuclear spin operations, reduce the overall fidelity, as well as demand lower thresholds to foster moderate spin photon entanglement generation rate. 

\section{Discussions}
In this work we showed an efficient nuclear spin - photon entanglement. 
We utilised NV center in diamond as a stable and long lived optically active quantum memory with either electron or nuclear spin serving as a memory qubit exceeding $100$ ms.
We used the active routing of the emmited zero-phonon line photons of the NV center to near deterministically convert the early late encoding of the photon qubit into dual rail mode suitable for measurement. 
We demonstrate all critical components of a quantum node within our work.
Our scheme achieves $\sim 70\%$ ($\sim60\%$) entanglement fidelity with electron (nuclear) spin photon states. 
This can be further improved by usage of nanophotonics and open cavity designs for efficient ZPL photon collection and more deterministic spin photon interaction. 
To this end, our approach could be further tested on G-IV defects in diamond, such as \cite{stas2022robust}, V2 centers in 4H-SiC \cite{fang2024experimental, hesselmeier2024high} and T centers in Silicon \cite{afzal2024distributed}. 
The nearby nuclear spin clusters, weakly coupled to the NV center \cite{vorobyov2022addressing} to this NV center, potentially open a possibility for creation of a robust and quantum repeater node added with error correction protocols.

\acknowledgments
We acknowledge support from the European Commission for the Quantum Technology Flagship project QIA (Grant agreements No. 101080128, and 101102140), the German ministry of education and research for the project QR.X (BMBF, Grant agreement No. 16KISQ013) and Baden-Württemberg Stiftung for the project SPOC (Grant agreement No. QT-6) as well as support from the project Spinning (BMBF, Grant agreement No. 13N16219) and the German Research Foundation (DFG, Grant agreement No. GRK2642).
We acknowledge Philipp Flad for generation of AR coating on a diamond chip. 
We acknowledge Sophie Hermans, Ronald Hanson, Regina Finsterhoelz and Guido Burkard for fruitful discussions and we thank QUBIG GmbH for providing the demultiplexer device and for fruitful collaboration within the QR.X project. 

%

\end{document}


\preprint{APS/123-QED}
\title{Supplementary material: Efficient nuclear spin - photon entanglement \\ with optical routing}
\author{Javid Javadzade}
 \affiliation{3rd Institute of Physics, IQST, and Research Centre SCoPE, University of Stuttgart, Stuttgart, Germany}  
\author{Majid Zahedian}
 \affiliation{3rd Institute of Physics, IQST, and Research Centre SCoPE, University of Stuttgart, Stuttgart, Germany}  
 \affiliation{Institute of Quantum Electronics, ETH Z\"urich, Switzerland}
\author{Florian Kaiser}
  \affiliation{Materials Research and Technology (MRT) Department, Luxembourg Institute of Science and Technology (LIST), 4422 Belvaux, Luxembourg}
  \affiliation{University of Luxembourg, 41 rue du Brill, L-4422 Belvaux, Luxembourg}
\author{Vadim Vorobyov}
\email[]{v.vorobyov@pi3.uni-stuttgart.de}
 \affiliation{3rd Institute of Physics, IQST, and Research Centre SCoPE, University of Stuttgart, Stuttgart, Germany}  
\author{J\"org Wrachtrup}
 \affiliation{3rd Institute of Physics, IQST, and Research Centre SCoPE, University of Stuttgart, Stuttgart, Germany}
 \affiliation{Max Planck Institute for solid state physics, Stuttgart, Germany}

                         
\maketitle
\onecolumngrid
\section{Cross-polarization setup for the ZPL photon collection}

\begin{figure} 
	\centering
	\includegraphics[scale=0.5]{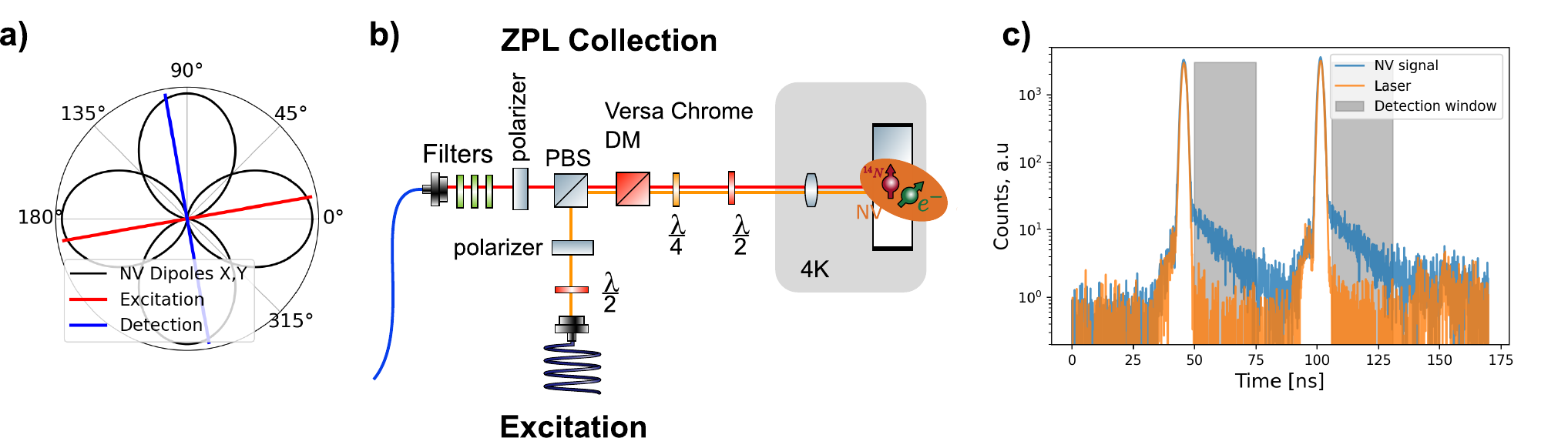}
	\caption{Cross Polarisation setup. \textbf{a)} The scheme of the optical dipole excitation of the $E_y$ transition with respect to the excitation (red) and detection (blue) polarization \textbf{b)} Optical setup scheme of the cross-polarization setup \textbf{c)} Collected photon statistics of arriving time for two consequetive optical pulse detected in the ZPL collection path, on (blue) and off (orange) the resonance of the Ey optical transition of the NV center.}
	\label{fig:x-pol}
\end{figure}

Zero-phonon line (ZPL) photons emitted from the NV center have approximately same frequency as the laser used for excitation. 
Consequently, we apply polarization filtering to separate the NV signal from the laser noise. 
We exploit the fact that the ZPL transitions used for generating entanglement($\ket{E_y}$) correspond to single dipole moment, and selection rules yield linearly polarized emission. 
Thus, cross-polarization scheme utilizes polarization of the excitation laser perpendicular to the polarization of the collected photons.

It operates as follows: a linearly polarized laser beam exits the optical fiber and first passes through a half-wave plate, which adjusts its polarization to match that of a polarizer. 
Then it hits a polarizing beam splitter, which reflects it toward the NV defect. 
Along the optical path, a quarter-wave plate compensates for the birefringence of the lens and the entire apparatus. 
Another half-wave plate is used to align the polarization of the incident light with the dipole moment of the NV center. 
By projecting the laser’s polarization onto the dipole axis, the transition is excited, resulting in an emitted photon with linear polarization that matches the dipole moment. 
Part of this linearly polarized light is able to pass through the polarization beam splitter and, after passing through an additional polarizer and optical filters, enters the polarization maintaining optical fiber. 
While this cross-polarization scheme is effective for isolating the NV signal, it has some drawbacks. 
The excitation polarization is not optimal because only the projection of the laser’s polarization onto the NV dipole axis can excite the transition. Additionally, some of the collected photons will be lost. 
However, these compromises are necessary to filter out the NV center's signal from the laser noise. 
We chose the $\lambda /2$ angle to increase the collected light from the NV center. 
By doing so, the excitation efficiency drops, so the higher laser power should be applied to excite the NV center. 
A special attention was applied to avoid excitation of the unwanted transitions due to the power broadening in the PLE spectrum. 
To evaluate the performance of the cross-polarization setup, we applied two consecutive optical $\pi-pulses$ and then measured the NV center's fluorescence, specifically assessing the lifetime of its excited state. 
The results are displayed in Figure \ref{fig:x-pol} c), which shows both the laser leakage through the cross-polarization scheme (orange curve) and the signal from the NV center (blue curve).
As observed, the cross-polarization setup does not entirely eliminate unwanted laser light. 
However, it is clear that the laser pulse and the NV signal are well separated in time. 
To address the remaining laser noise, we implemented "time filtering." 
This approach involves considering only the photons that arrive at the detector within a specified time window (indicated by the grey area in Figure  \ref{fig:x-pol} c). This additional step helps to isolate the NV center’s signal more effectively despite the incomplete suppression of the laser light by the cross-polarization setup.
\section{Spectral diffusion and charge resonance check}
\begin{figure} 
	\centering
	\includegraphics[scale=0.5]{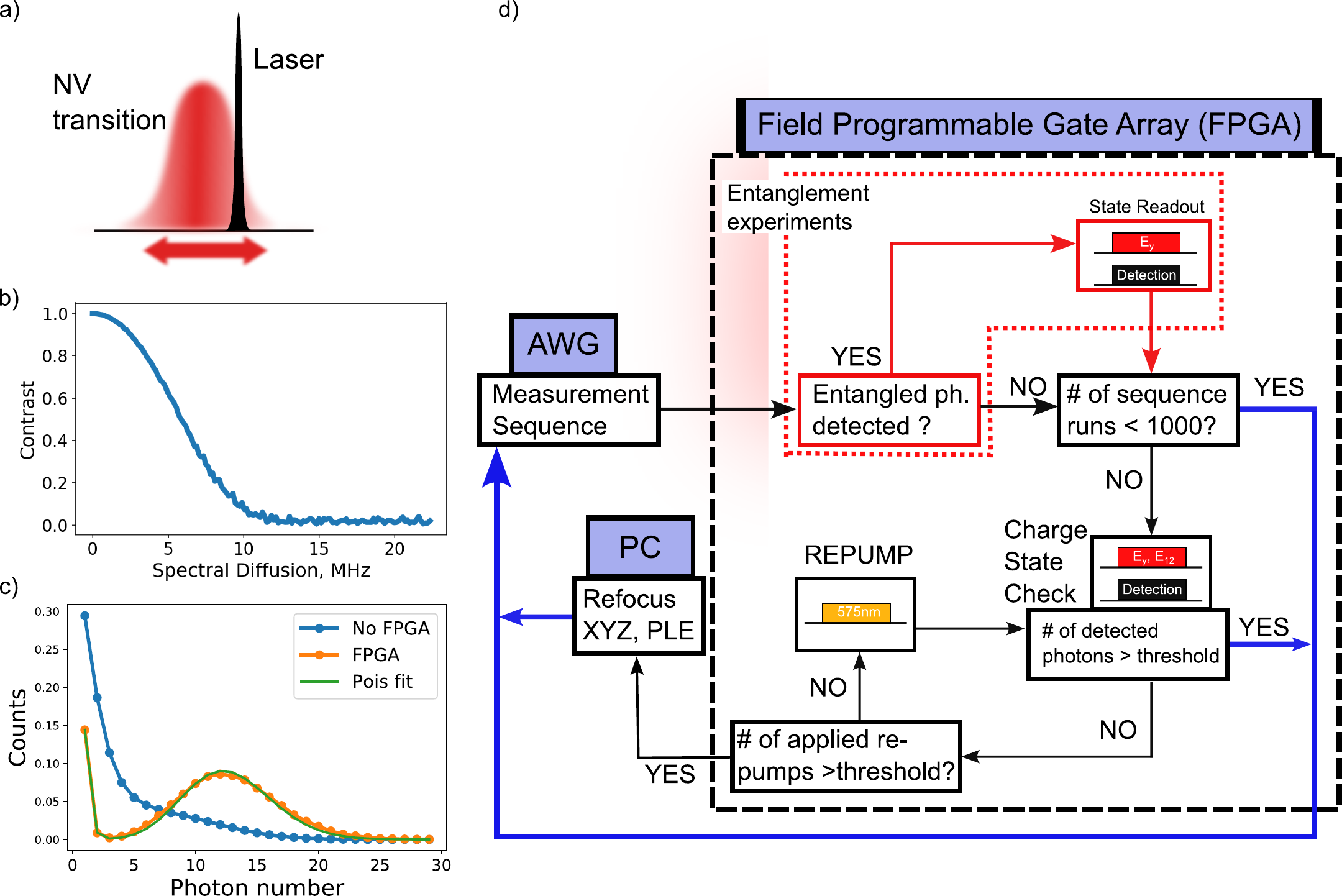}
	\caption{\textbf{a)} Scheme of spectral diffusion \textbf{b)} Monte-Carlo simulation of the spin-photon entanglement contrast decay with spectral diffusion intensity RMS \textbf{c)} Photon statistics histogram before (blue) and after CRC filtering (orange curve). Green is an ideal Poisson distribution \textbf{d)} Block diagram scheme of the protocol of the spin photon entanglement sequence}
	\label{fig:FPGA}
\end{figure}
Fluctuating charge environment creates electrical field drift at the position of the NV center, which leads to a shift its energy levels. 
Thus, the transition frequency moves away from the exciting laser frequency (Figure \ref{fig:FPGA} a). 
Due to the power broadening, one is able to efficiently excite the NV center.
The optical path is phase stabilised with the means of excitation laser frequency.
The emitted photon of a different frequency will deviate in acquired phase and when averaging the data, the total contrast of spin-photon oscillations will be reduced (Figure \ref{fig:FPGA} b). 
To avoid this we used Field Programmable Gate Arrays (FPGA), which was used to check, that the NV center is in the correct charge state ($NV^-$) and is on resonance with the laser (Figure \ref{fig:FPGA} d).
The measurement sequence runs $1000$ times and after this, Charge Resonance Check (CRC) routine happens - 2 laser pulses $E_y$ and $E_{1,2}$ (specific to spin projections $0$ and $\pm1$) are sent to NV center simultaneously and amount of detected photons is counted. 
If the value surpasses the threshold, the NV center is prepared to initiate the experiment. 
If it falls below the threshold, it could indicate either an improper charge state of the $NV$ center or a lack of resonance with the lasers. 
In either scenario, we employ a weak $575$ nm laser resonant to $NV^0$ pulse to perturb the charge bath around the NV center. 
This process is repeated until the NV center achieves resonance with the lasers and the measurement sequence can be continued, or until pre-defined number of recharging attempts followed by a PLE scan and refocus of the confocal objective to NV center. 
Moreover, in entanglement experiments, we employ FPGA to speed up the data acquisition  process. 
Detecting an entangled photon is a low-probability event due to the low Debye-Waller factor and limited collection efficiency of our setup.
Therefore, it is efficient to read out the electron spin state only when the corresponding resonant  ZPL photon has been detected (Figure \ref{fig:FPGA} d, red shaded area). This approach conserves time by avoiding unnecessary state readouts.
The resonance position of the NV center experiences spectral diffusion. 
When performing multiple Charge Resonance Checks by counting photons during illumination with two resonant lasers, the resulting photon distribution is a sum of Poisson distributions corresponding to different detuning values (spectral diffusion). 
Using FPGA based CRCs narrows down the acceptable range of detunings. 
This can be best seen in (Figure \ref{fig:FPGA} c), where a photon statistics with and without CRC shows reduction of count fluctuations, originating from large spectral diffusion as well as charge state switching. 
Therefore, using FPGA is crucial for suppressing spectral diffusion and accelerating the experiment.
\section{Time-bin to spatial modes conversion}

\begin{figure} 
	\centering
	\includegraphics[scale=0.5]{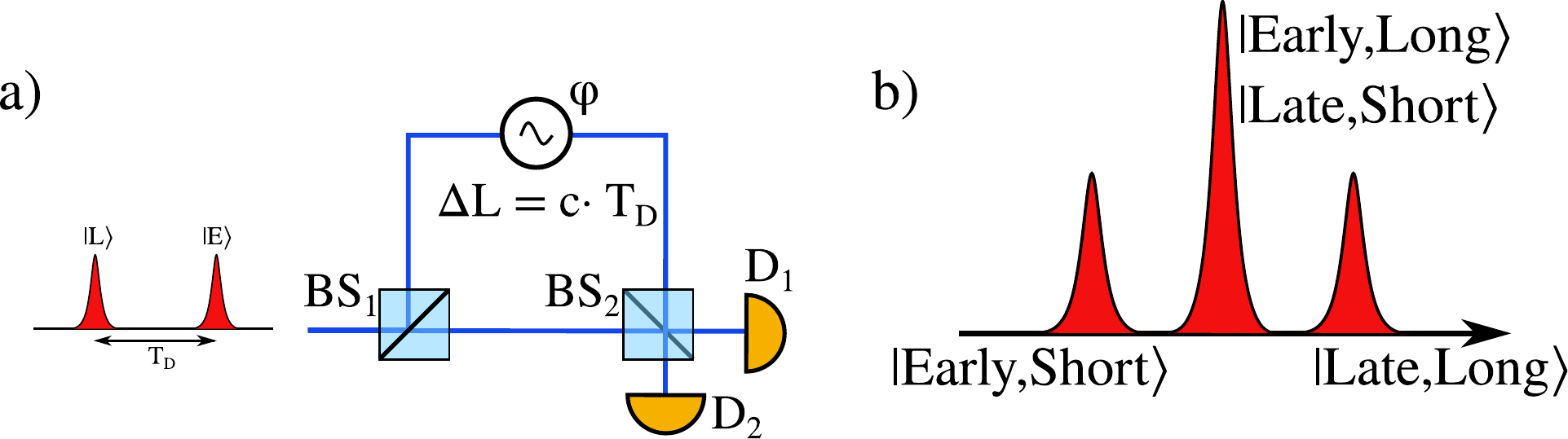}
	\caption{Interferometer}
	\label{fig:Interferometer}
\end{figure}

Usually the unbalanced Mach-Zehnder interferometer is used to control the phase of the photon qubit in time-bin encoding. The typical interferometer has a 50/50 Beam Splitter instead of EOD (Figure \ref{fig:Interferometer} a) so that, the photon routing is probabilistic. If one plots histogram of photons arrival time to the detectors - the 3 peaks must be seen:
\begin{enumerate}
\item $\ket{Early}$ photon takes $\ket{Short}$ path\\The left (earliest) peak on the Figure \ref{fig:Interferometer} b
\item $\ket{Late}$ photon takes $\ket{Long}$ path\\The right (latest) peak on the Figure \ref{fig:Interferometer} b
\item $\ket{Early}$ photon takes $\ket{Long}$ path and the $\ket{Late}$ photon takes $\ket{Short}$ path\\The central peak  on the Figure \ref{fig:Interferometer}, particularly in this peak the information "In which time bin the photon was emitted" is erased
\end{enumerate}
In this configuration only the central peak could be used, so half of the photons (corresponding to the left and to the right peak) are discarded. The Electro-Optical Deflector particularly improves this scenario, by guiding the $\ket{Early}$ coming photon in the $\ket{Long}$ arm of the interferometer and $\ket{Late}$ one to the $\ket{Short}$ arm. If the delay between RF drive of EOD and optical pulses synchronized correctly, then 2 undesired peaks must be suppressed on the histogram (Figure \ref{fig:Interferometer} b or Figure 3c of the main text).

\section{Spin photon entangled state}
In this section we will analyze the state of the spin-photon system at different steps of the experiment. We started with a state Eq. \ref{eq:psi1} (see main text Eq. 1), which was sent to the unbalanced Mach-Zehnder Interferometer:
\begin{equation}
\ket{\psi_1} = \frac{\ket{0,Late}+\ket{1,Early}}{\sqrt{2}}
\label{eq:psi1}
\end{equation}
After the EOD, the $\ket{Early}$ photon will be guided to the $\ket{Long}$ arm and the $\ket{Late}$ photon to the $\ket{Short}$ one, so transforming the system state into:
\begin{equation}
\ket{\psi_2} = \frac{\ket{0,Late,Short}+\ket{1,Early,Long}}{\sqrt{2}}
\label{eq:psi2}
\end{equation}
from now on, we can use labels, corresponding to spatial modes instead of temporal, because they have one-to-one correspondence and after the interferometer the two temporal modes won't be distinguished anyway. The fiber strecher introduces relative phase $\phi$ between $\ket{Long}$ and $\ket{Short}$ modes, so the state of the system just before the second interferometer's Beam Splitter is (also see Eq. 3 in the main text):
\begin{equation}
\ket{\psi_3} = \frac{\ket{0,Short}+e^{i \phi}\ket{1,Long}}{\sqrt{2}}
\label{eq:psi3}
\end{equation}
After the second Beam Splitter, the photon can take either the path to one detector or to another with 50\% chances, giving the state:
\begin{equation}
\ket{\psi_4} = \frac{\ket{0,D_1}+\ket{0,D_2}+e^{i \phi}\ket{1,D_1}-e^{i \phi}\ket{1,D_2}}{2}
\label{eq:psi4}
\end{equation}
Where $D_1$ and $D_2$ - spatial modes, corresponding to two different detectors. The negative sign in the last term appears due to the properties of the quantum beam splitter operator yielding   $\ket{\psi_4}$:
\begin{equation*}
\ket{\psi_4} = \frac{1}{2} \ket{D_1} \otimes (\ket{0}+e^{i \phi}\ket{1})+\frac{1}{2} \ket{D_2} \otimes (\ket{0}-e^{i \phi}\ket{1})
\label{eq:psi4_2}
\end{equation*}
which heralds the photon click on detector number 1, $\ket{D_1}$ - with the state of the $NV$ center being $\frac{1}{\sqrt{2}}(\ket{0} + e^{i\phi}\ket{1})$, and click of $D_2$ to $\frac{1}{\sqrt{2}}(\ket{0} - e^{i\phi}\ket{1})$ state of the NV. One can show, that $\pi /2$ rotation applied to these states gives the probabilities shown in the Main text, Equation 4.

\section{14N initialization} 
\begin{figure} 
	\centering
	\includegraphics[scale=0.5]{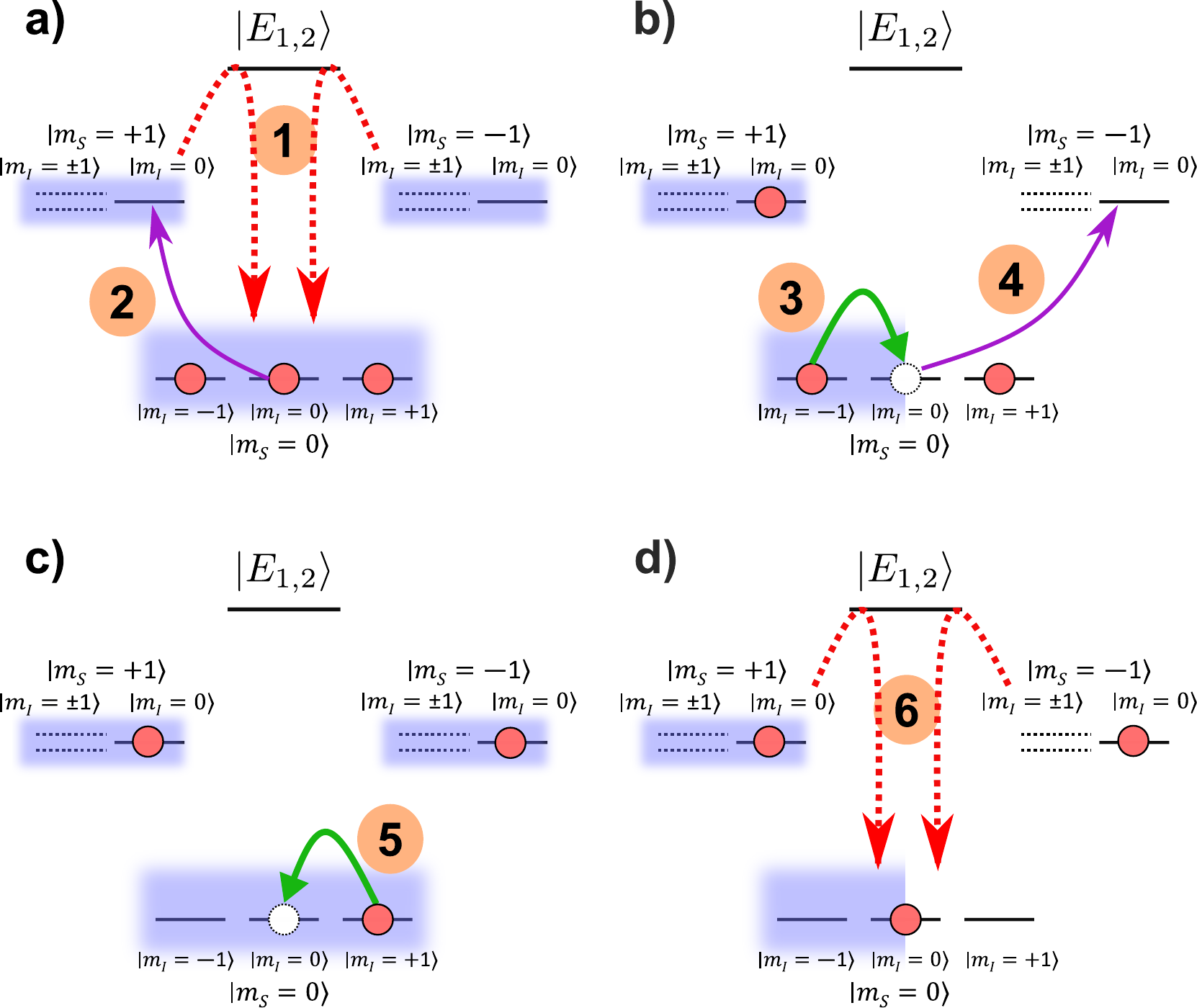}
	\caption{Nuclear spin initialisation steps a), b), c), d) Optical reinitialisation with $E_1,2$ laser into $m_s = 0$ $m_i = 0$}
	\label{fig:14N_init}
\end{figure}

We assume the Nuclear spin to be initially in a fully mixed thermal state. 
Hence, the protocol of Nuclear spin - photon entanglement starts with $^{14}N$ initialization. 
Due to lack of direct access to the nuclear spin, the initialization happens through electron spin and takes following steps depicted in Figure \ref{fig:14N_init}:

\begin{enumerate}[label=\protect\circled{\arabic*}]
\item The protocol starts with electron spin initialisation by driving $E_{1,2}$ transition (selective to $\ket{m_s=\pm 1}$)
\item As a second step the Controlled NOT gate applied between electron levels $\ket{m_s = 0} \rightarrow \ket{m_s = +1}$ to store the nuclear $\ket{m_I = 0}$ population in another electron spin projection subspace
\item Then via RF drive between $^{14}N$ hyperfine sublevels in $\ket{m_s = 0}$ subspace, the population transfer happens $\ket{m_I = +1} \rightarrow \ket{m_I = 0}$
\item Following CNOT stores the $\ket{m_I = 0}$ projection into $\ket{m_s = -1}$ electron subspace
\item RF pulse transfer the remaining $\ket{m_I = -1}$ population to $\ket{m_I = 0}$
\item As on the first step, the $E_{1,2}$ laser pulse initialize the electron spin without significantly affecting nuclear
\end{enumerate}

